\documentclass[journal=jacsat,manuscript=article]{achemso}

\usepackage[version=3]{mhchem} 
\usepackage{graphicx}  
\usepackage[caption=false]{subfig} 

\author{Jia-Yang Chen}
\altaffiliation{Department of Physics and Engineering Physics, Stevens Institute of Technology, Hoboken, New Jersey 07030, USA}
\alsoaffiliation{Center for Quantum Science and Engineering, Stevens Institute of Technology, 1 Castle Point Terrace, Hoboken, New Jersey  07030, USA}

\author{Yong Meng Sua}
\altaffiliation{Department of Physics and Engineering Physics, Stevens Institute of Technology, Hoboken, New Jersey 07030, USA}
\alsoaffiliation{Center for Quantum Science and Engineering, Stevens Institute of Technology, 1 Castle Point Terrace, Hoboken, New Jersey  07030, USA}

\author{Heng Fan}
\altaffiliation{Department of Physics and Engineering Physics, Stevens Institute of Technology, Hoboken, New Jersey 07030, USA}
\alsoaffiliation{Center for Quantum Science and Engineering, Stevens Institute of Technology, 1 Castle Point Terrace, Hoboken, New Jersey  07030, USA}

 \author{Yu-Ping Huang}
 \altaffiliation{Department of Physics and Engineering Physics, Stevens Institute of Technology, Hoboken, New Jersey 07030, USA}
\alsoaffiliation{Center for Quantum Science and Engineering, Stevens Institute of Technology, 1 Castle Point Terrace, Hoboken, New Jersey  07030, USA}
 \email{yhuang5@stevens.edu}

\title[An \textsf{achemso} demo]
  {Naturally Phase Matched Lithium Niobate Nanocircuits for Integrated Nonlinear Photonics }

\abbreviations{IR,NMR,UV}

\begin{document}







\begin{abstract}
  High complexity, dense integrated nanophotonic circuits possessing strong nonlinearities are desirable for a breadth of applications in classical and quantum optics. In this work, we study natural phase matching via modal engineering in lithium niobate (LN) waveguides and microring resonators on chip for second harmonic generation (SHG). By carefully engineering the geometry dispersion, we observe a $26\%~W^{-1}cm^{-2}$ normalized efficiency for SHG in a waveguide with submicron transverse mode confinement. With similar cross-sectional dimensions, we demonstrate phase matched SHG in a microring resonator with 10 times enhancement on the out-coupled second-harmonic power. Our platform is capable of harnessing the strongest optical nonlinear and electro-optic effects in LN on chip with unrestricted planar circuit layouts. It offers opportunities for dense and scalable integration of efficient photonic devices with low loss and high nonlinearity.
\end{abstract}

\section{Introduction}
The discovery of optical harmonic generation by Franken et al \cite{FrankenSHG}, where ultraviolet (UV) light was created from a red ruby laser in a single-crystal quartz, embarks the era of nonlinear optics. Among different materials, lithium niobate is widely sought-after due to its large  second-order nonlinear susceptibilities (e.g., $\sim$27 pm/V), wide transparency window across UV and mid-infrared (0.35--5.2 $\mu$m), and ultrafast electro-optic responses. With those exceptional attributes, optical devices based on bulk LN have excelled in a wide range of applications, such as electro-optic modulators \cite{doi:10.1063/1.1906311}, nonlinear wavelength mixers and converters \cite{Shentu:13}, and non-classical light sources \cite{Kaiser:16}.

Conventionally, LN waveguides have been fabricated using metal diffusion \cite{Zhang2015}, ion implantation \cite{Montanari:12}, proton exchange \cite{Cai:15}, and mechanical dicing \cite{thinfilmdicingVolk:16}. However, they suffer poor mode confinement and large bending radii, as limited by either the low refractive index contrast or bad dicing resolution. More recently, there have been growing interest and efforts in incorporating LN into integrated photonics \cite{doi:10.1021/acsphotonics.5b00126, Wang:18,Jiang2016, zenojyc}. To this end, efficient and scalable circuits with low loss are essential for exploiting the strong nonlinearities with LN for broad applications in both classical and quantum domains, including communications \cite{Wang:18}, sensing \cite{Jiang2016}, all-optical signal processing \cite{5546889}, spectroscopy \cite{LiNiospdcspectroscopy}, and quantum information science \cite{zenojyc}. However, LN nanostructures are difficult to fabricate with low loss, due to its inert chemical properties and physical hardness \cite{ULLIAC20161,huhuiplasmaetch}.

Due to those difficulties, research interests in the pursuit of second-order nonlinear nanophotonics have shifted to other materials that are easier to integrate, such as AlN \cite{Guo:16}, GaAs \cite{doi:10.1021/ph500054u}, InP \cite{InpAPL2010}, and BaTio3 \cite{lowlossBaTio}. However, the success has been limited as none of those materials could match the excellent optical properties of LN. Another approach is to utilize heterogeneous structures with the presence of low loss SiN nanowires to aid mode confinement where lightwaves are guided in a etch-free LN thin film for nonlinear interaction \cite{Chang:16}. This is achieved by first bonding the LN film onto a wafer substrate via ion slicing and then sputtering SiN on top the film, where nanostructures are etched to guide and confine light without etching the LN film itself \citep{Chang:16}. Yet, progressed are awaited to improve the mode confinement while reducing the propagation loss. 

Very recently, there have been encouraging progress in fabricating nanophotonic waveguides and microrings with high index contrast and low loss  based on a lithium niobate-on-insulator (LNOI) platform \cite{Zhang:17,Krasnokutska:18}. Compared with bulk LN devices, they offer benefits such as much reduced footprint, low cost per device, and significant field enhancement, which are desirable for high complexity and dense integrated photonics \cite{BOWERSlnoi}. Furthermore, integrated monolithic LNOI circuits are inherently stable and can avoid coupling losses during device serialization. They can be an excellent candidate for quantum nonlinear optics, especially for quantum information processing not easily realizable or feasible in other existing integrated platforms \cite{Caspani2017,Samkharadzeeaar4054}. In this pursuit, Wang et al. \cite{Wang:17} demonstrated naturally phase matching on X-cut LNOI wafer, with the normalized efficiency $\sim 41 \%W^{-1}cm^{-2}$  for second-harmonic generation (SHG). Yet, the phase matching is restricted to only a particular direction of light propagation, as the birefringence of the LN material will introduce phase mismatch otherwise. This constitutes a constraint on the realizable photonic circuits, especially those requiring serialization of dislike elements for complex functionalities. 

To address this constraint, here we demonstrate modal phase matching on a monolithic Z-cut LNOI, thereby achieving SHG in submicron LN waveguide with $\sim 26 \%W^{-1}cm^{-2}$ efficiency, and in microring resonator with over 10 times enhancement. The technique involves compensating the chromatic dispersion of the LN material by precisely engineering the geometric dispersion of the optical transverse modes. It can realize natural phase matching between far-detuned wavelengths, in the current example between 1550 nm and 775 nm, both in the quasi-transverse-magnetic modes. This allows us to fully utilize d$_{33}$, the by-far largest tensor element of LN's second-order nonlinear susceptibility, along any light propagation direction on chip. For the current Z-cut LNOI, it enables arbitrary layouts of nonlinear optical circuits--such as a spiral waveguide for optical delay followed by a microring for quantum frequency conversion--on the XY plane with the highest efficiency and natural phase matching.    

 

\section{Results and discussion}
\subsection{Simulation and fabrication}

To achieve natural phase matching for SHG of telecom light in LN nanowaveguides, we need to compensate for the large chromatic dispersion in the LN material. The first solution is by utilizing the birefringence and geometric dispersion to offset the chromatic dispersion between the pump and SH wavelengths \cite{moore2016efficient,zenojyc}. The phase matching can be obtained by equating the effective refractive indexes (n$_\mathrm{eff}$) of the fundamental quasi-transverse-electric (quasi-TE) mode and fundamental quasi-transverse-magnetic (quasi-TM) mode, each for the pump and SH lightwaves. The primary disadvantage of this approach, however, is that the $\chi^{(2)}$ nonlinear parametric processes are through the $d_{31}$ and $d_{32}$ tensor elements, where the nonlinear coefficients are 4 to 5 times lower than $d_{33}$. Another solution is by using higher-order modes for SH light, together with geometric dispersion, to compensate for the chromatic dispersion \cite{Guo:16,Wang:17}. We can match the effective mode indexes of the interacting pump and SH wavelengths (both in quasi-TE or quasi-TM polarization) with precisely fabricated waveguide cross-section, for achieving modal phase matching. Comparing with the first approach, it allows us to exploit the largest nonlinear coefficient of LN, the $d_{33}$ tensor element which can be 27 pm/V or even higher. Such modal phase matching can be applied to LNOI with X-cut or Z-cut crystal orientation. For phase matching between quasi-TE modes on X-cut wafers, the propagation direction of the interacting lightwaves is restricted to only along the Y-axis, because otherwise the birefringence will introduce large wavevector mismatch to destroy the natural phase matching \cite{Wang:17}. In contrast, Z-cut LNOI wafers provide access to the $d_{33}$ tensor in all propagation directions of light in quasi-TM modes, thus capable of maintaining phase matching on the entire X-Y plane for arbitrarily-guided structures, such as microrings and microdisks. The same design allows to access the largest $r_{33}$ electro-optic coefficient of LN with modulation induced by an applied electric field along the Z axis \cite{Chen:13}.

\begin{figure}[htbp]
\centering
  \subfloat{
   \includegraphics[width=0.43\textwidth]{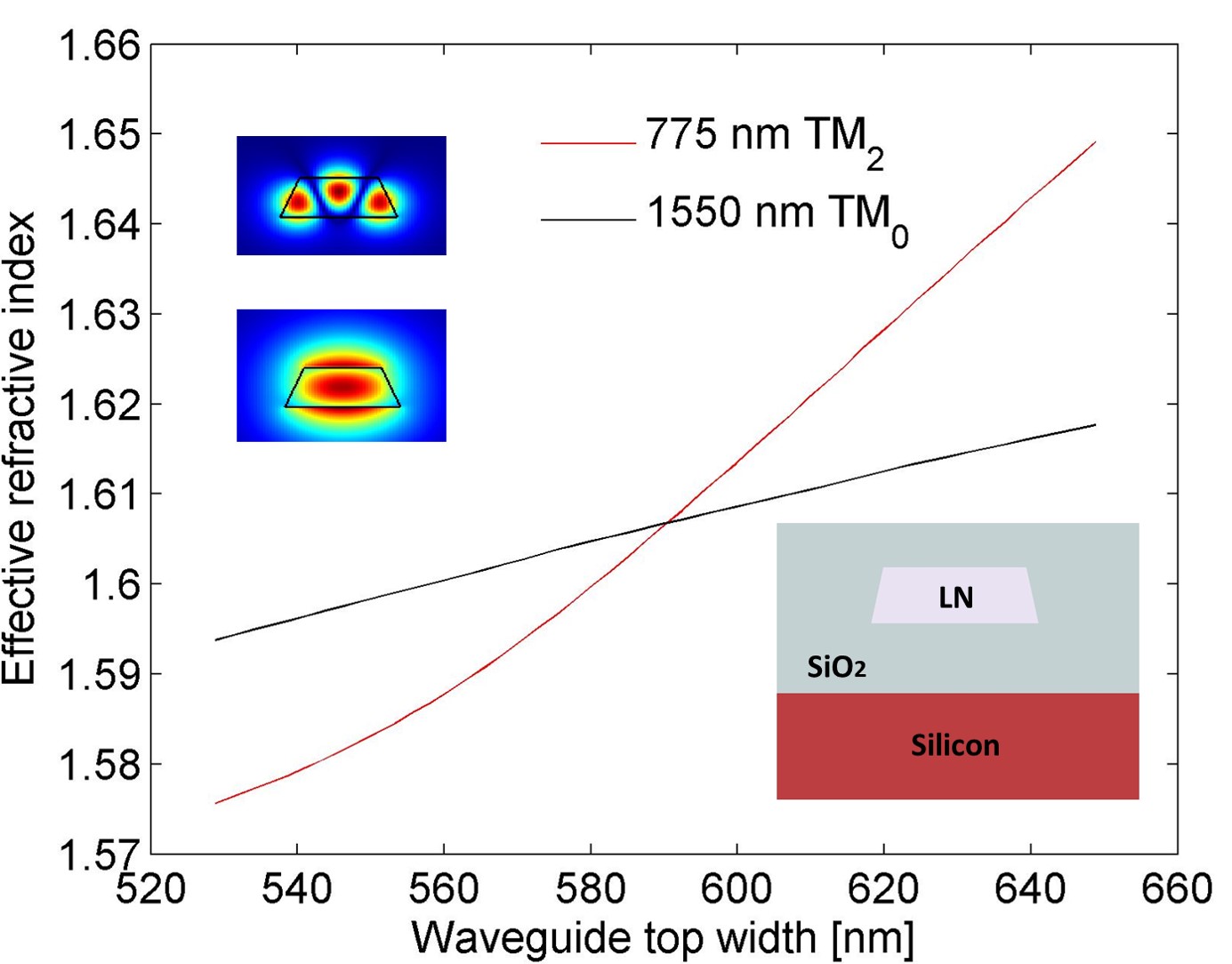}
   \label{fig:sub11}
 }
 \qquad
 \subfloat{
   \includegraphics[width=0.45\textwidth]{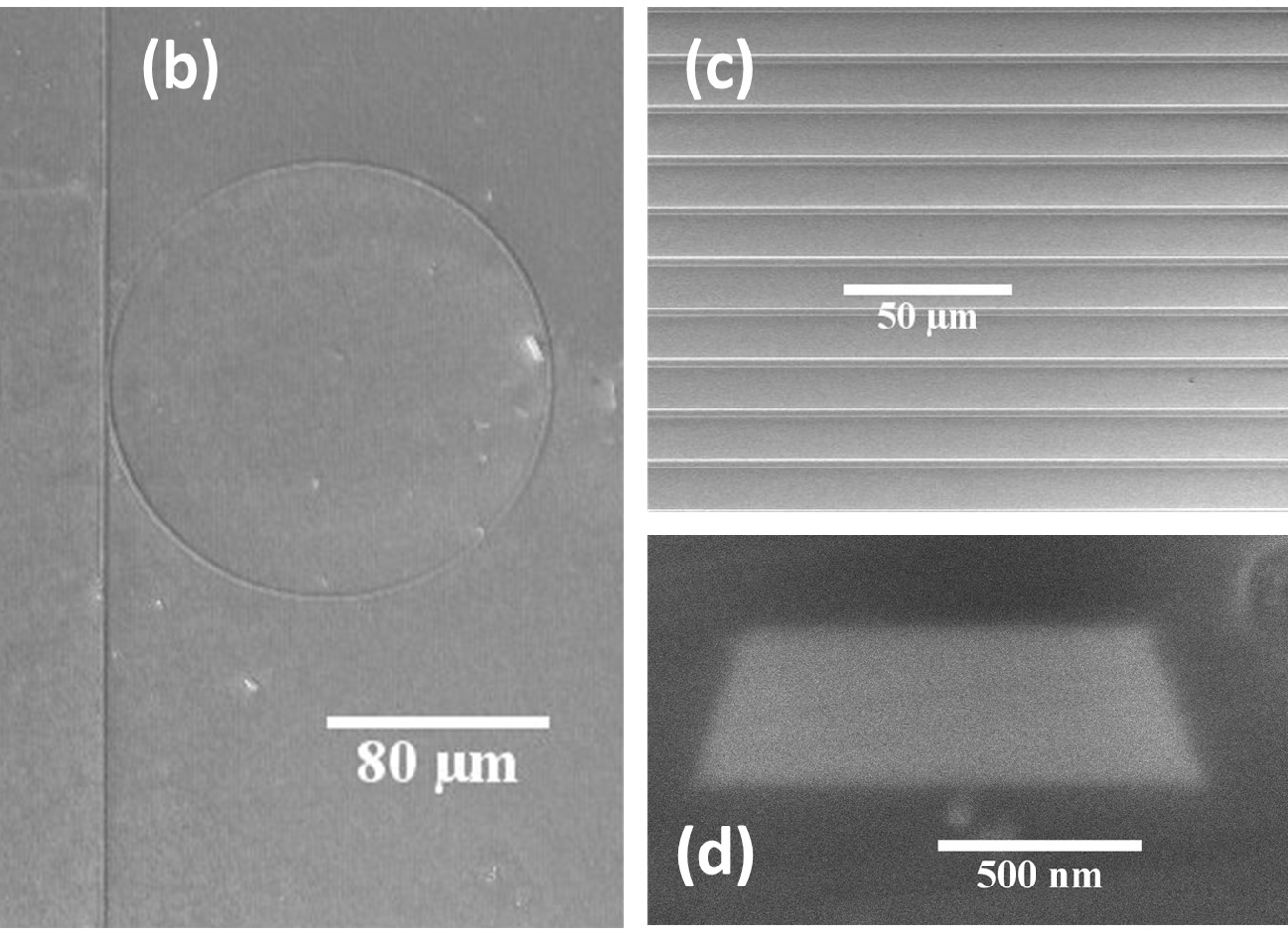}
   \label{fig:sub12}
} 
 \caption{(a) Phase matching curve of nanophotonic LN waveguide. Inset are the mode profiles and waveguide schematic. SEM images of a microring resonator (b), a waveguide (c), and the cross-section of a typical waveguide (d).}
 \label{figure1}
\end{figure}

To explore those advantages, we study modal phase matching in both straight waveguides and microring resonators, both on Z-cut LNOI. 
For the actual cross-section geometry of the fabricated waveguides, we simulate the transverse profiles and refractive indexes of the TM$_0$ modes at the pump wavelength and the TM$_2$ modes at the SH wavelength, as depicted in Fig.~\ref{figure1}(a). It is shown that natural phase matching between 1550-nm TM$_0$ and 775-nm TM$_2$ modes can be obtained by varying the top width of the LN waveguide around 590 nm. The two insets on the top left in Fig.~\ref{figure1}(a) shows the simulated profiles with 590 nm top width and $70^\circ$ sidewall angle, for which natural phase matching occurs. For the microrings, the phase matching is achieved similarly with small contribution from the ring bending. The inset on the bottom right illustrates the cross-section of the actual LNOI waveguide. The refractive index contrast $\Delta n$ between LN ($\sim$ 2.23) and the silicon dioxide cladding ($\sim$ 1.45) is about $0.6$, which gives rise to negligible bending loss even for small radii down to 15 $\mu$m, thus providing a promising route for dense devices integration and strong mode confinement. 

The waveguide and microring resonator are fabricated on a LNOI (by NANOLN Inc.), which is a 400 nm thick LN thin film bonded on a 2 $\mu$m silicon dioxide layer above a silicon substrate. We use FOX-16 (flowable oxide, Down Corning) electron-beam resist and define the pattern of the desired structures by using electron-beam lithography (Elionix ELS-G100, 100 keV). After development, an optimized ion milling process (AJA e-beam evaporate system) is used to etch the structure with smooth sidewalls and an optimum sidewall angle ($67^\circ \pm 3^\circ$) as shown in Fig.~\ref{figure1}(d). RCA I (5:1:1, deionized water, ammonia and hydrogen peroxide) solution is then used to remove the re-sputtered material during the etching process. Next, a layer of 2 $\mu$m thick silicon dioxide is deposited via plasma enhanced chemical vapor deposition (PECVD) as over-cladding, which also serves as a protection layer for the dicing and polishing procedures later on. 

In practice, it is challenging to fabricate the LN structures with desirable geometries for phase matching, especially due to the requirement of accurate waveguide dimensions. The difficulties arise from the non-uniformity ($\pm$ 20 nm) of the LN layer across the whole wafer, the local refractive index variation, and the accumulation of fabricate errors ($\pm$ 30 nm). Once the optimum waveguide width is determined by using a finite-element-method (FEM) simulation tool, we define the pattern for waveguides with various width ranging from 500 nm to 700 nm with 2 nm step size. Figure~\ref{figure1}(c) shows the fabricated waveguide array, where each waveguide is characterized to identify the optimally phase matching and compared with the simulation results. This task is crucial for us to understand the limitation and accuracy of our fabrication process, thus providing valuable guidance to proceed with the microring structure. 

\begin{figure}[htbp]
\includegraphics[width=5in]{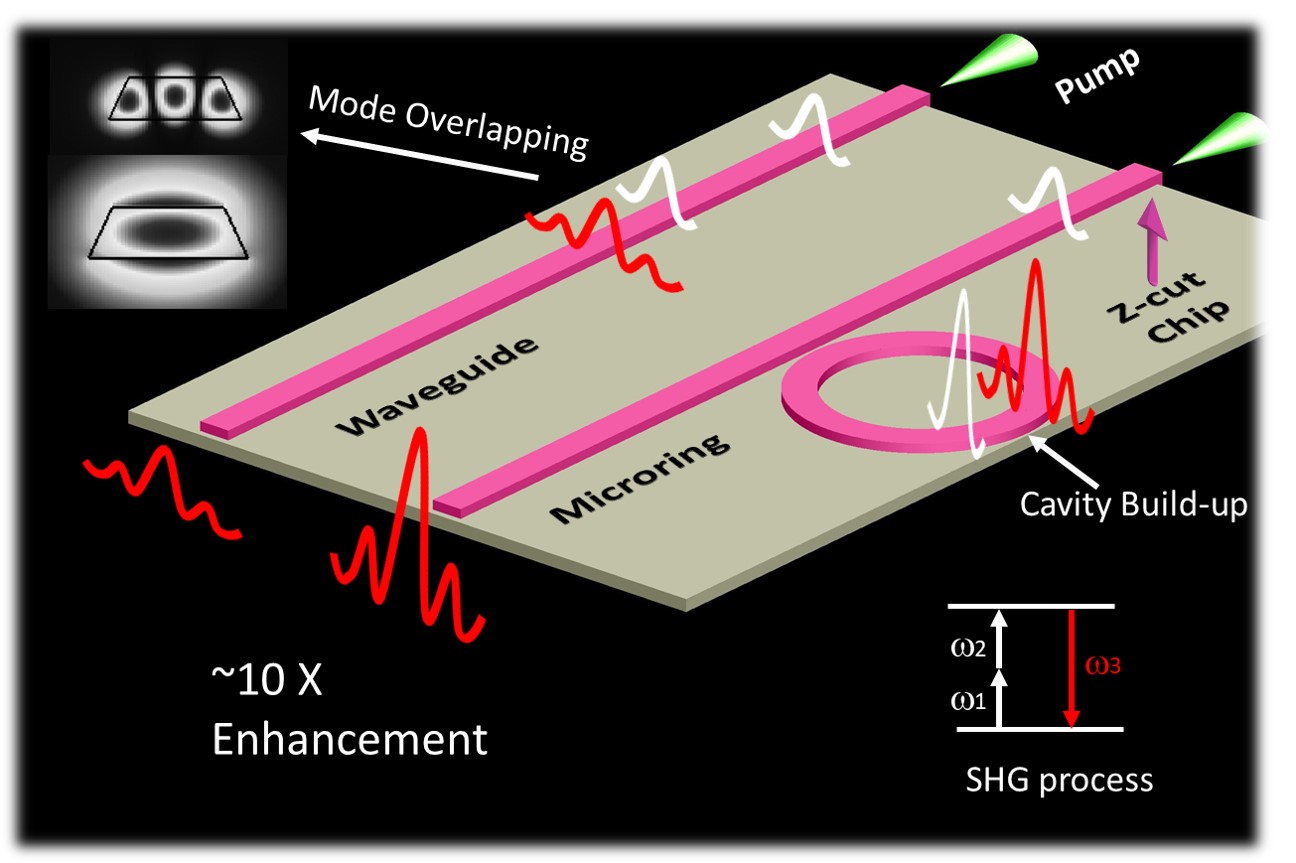} 
  \centering
\caption{Schematic of second harmonic generation in a waveguide and microring resonator. The inset in the left corner represents mode overlapping between 1550-nm TM$_0$ and 775-nm TM$_2$ modes. The inset in the right corner is the energy scheme of nonlinear frequency conversion process, which is SHG when $\omega_1=\omega_2$.}
\label{schematic}
\end{figure}

\subsection{Experiment}
\label{experiment}
For the SHG measurement, we use a lensed fiber to couple the pump light at 1550 nm (coupling efficiency $\approx$ 15$\%$) into the waveguide and collect the generated SH light by using another lensed fiber \cite{Guo:16}. This coupling loss can be reduced to 1 dB per facet by inversely tapered waveguides \cite{inversetaper}. To quantify the SHG efficiency, we measure both the spectrum and optical power of the SH light by using an optical spectrum analyzer (OSA) and a power meter, respectively. However, the coupling efficiency for the SH light is quite low ($\leq$ 10$\%$), as it is in the TM$_2$ mode and the lensed fiber is not optimized for this wavelength. An accurate estimation of the output coupling efficiency for the SH light is difficult, which might lead to errors in calculating the nonlinear conversion efficiency. Hence, we  also perform sum frequency generation (SFG) to validate the SHG efficiency determined through the power measurement. For this purpose, the SFG is carried out between two nearby wavelengths centered around 1548 nm, so that the phase matching condition remains nearly the same with SHG. Then the conversion efficiency is derived from the signal depletion at the waveguide output by using a high speed InGaAs photodetector and a 20 GHz oscilloscope. The experimental setup is shown in S.1 (see supplementary for details). The signal depletion measurement is more accurate in determining the nonlinear conversion efficiency, as it is a direct measurement requiring only the coupling efficiency for the infra-red (IR) light in the TM$_0$ mode \cite{Allgaier2017}. This is in contrast to the method by measuring the SH power, where the coupling losses for both the pump and SH light need to be measured accurately.

\subsubsection{Waveguide}
\begin{figure}[htbp]
\centering
 \subfloat{
   \includegraphics[width=0.45\textwidth]{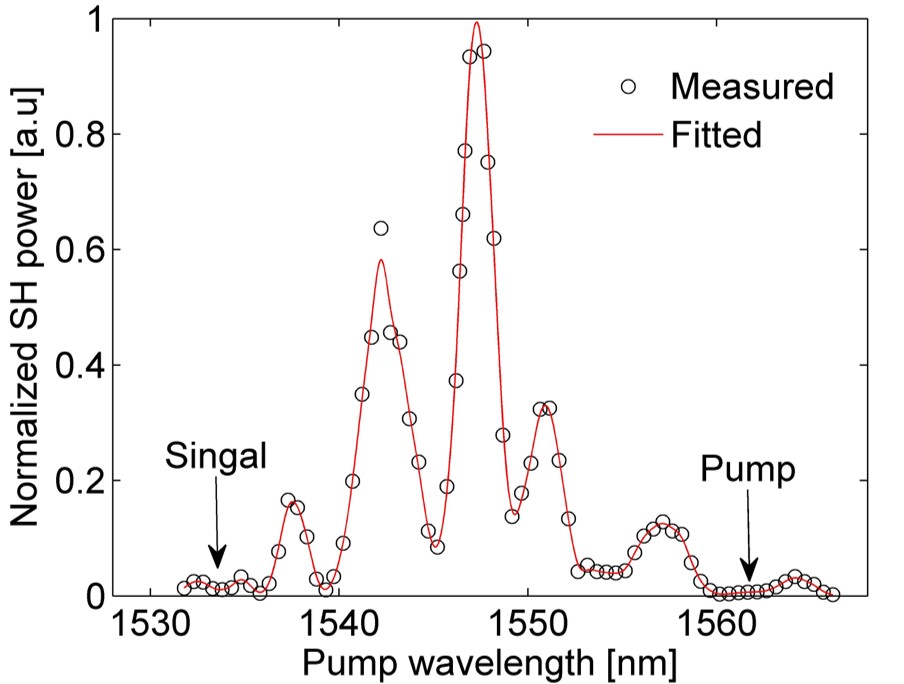}
   \label{fig:sub21}
 }
 \qquad
 \subfloat{
   \includegraphics[width=0.46\textwidth]{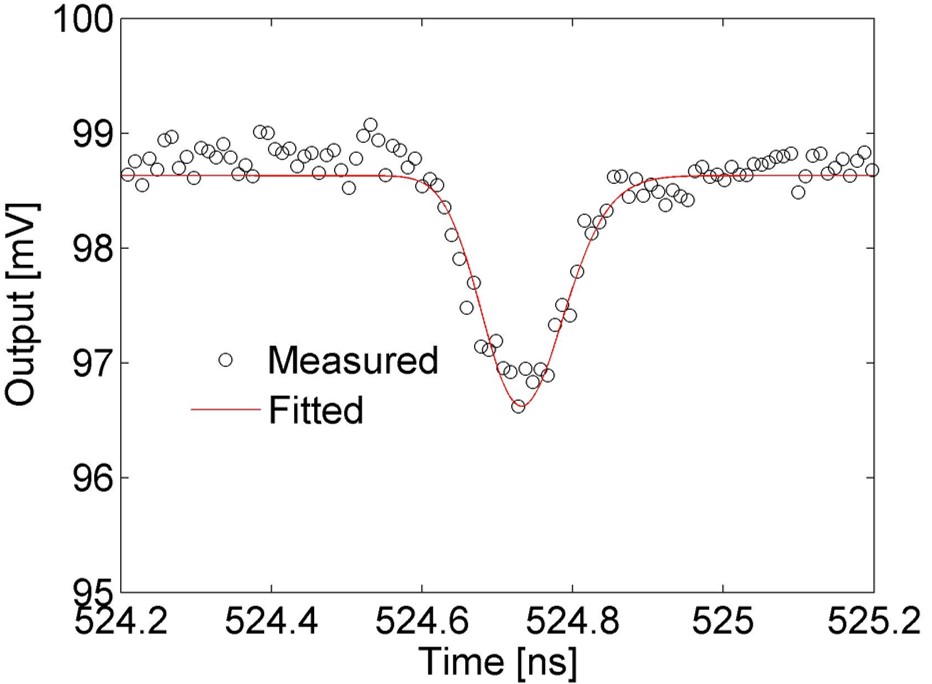}
   \label{fig:sub22}
 }
 \caption{(a) Phase matching curve in IR band for the waveguide with L = 1 mm, top width = 594 nm and sidewall = 70 $^\circ$. (b) Sum frequency generation on the same waveguide. The pump and signal wavelength for SFG are  around 1562 nm and 1534 nm.}
 \label{figure3}
\end{figure}

We measure the SHG phase matching bandwidth of the LN waveguide by sweeping a continuous-wave (CW) pump laser across telecom C-band in the non-depleted pump regime and recording the resulting SH power. As depicted in Fig.~\ref{figure3}(a), we observe multiple side peaks in addition to a central phase matching peak for a 1-mm long waveguide. The 3-dB bandwidth of the central peak in terms of the pump wavelength is measured to be about 1.9 nm. However, our simulation based on the measured waveguide cross-section predicts a 0.9 nm bandwidth. This disagreement could come from several factors such as thickness variation and compositional inhomogeneity of the LN thin film, as well as fabrication errors during the extended electron beam writing over 1 mm and ion milling dry etching (see supplementary for details). 

We next study the thermo-optic and thermal expansion effects on the phase matching. For this purpose, we measure the temperature dependency of both the phase matching wavelength and the corresponding SH efficiency. The central wavelength of the main phase matching peak is found to shift by $+$0.065 nm per 1$^{\circ}$C increase, which is consistent with our theoretical simulation results (see supplementary for details).

For accurate SFG measurement, we choose the pump and signal wavelengths that are optimally phase matched for SFG while sufficiently phase mismatched for each's own SHG, based on the results in Fig.~\ref{figure3}(a). After optimizing the polarization and temporally aligning both pump and signal pulses, we observe a clear $2\%$ dip ($\eta_{dip}$, averaged over 50 million measurements) on the signal with 7.7-watt pump peak power on chip. Considering the actual waveguide length of 1 mm, the normalized conversion efficiency is then $\eta = \frac{\eta_{dip}}{P_{pump}L^2} = 26\%W^{-1}cm^{-2}$, which is 2.3 times less than the predicted value of $ 61 \%W^{-1}cm^{-2}$ (see supplementary material). This discrepancy and the non-ideal phase matching curve is likely caused by the LN thin film inhomogeneity and fabrication errors accumulated over the 1 mm waveguide span, which can include stitching and variation errors. The same factors also lead to the multiple side peaks on both sides of the primary peak as shown in the phase matching curve in Fig.~\ref{figure3}(a). In practice, the fabrication errors can be mitigated by using a  microring structure to effectively extend the interaction length within a rather small footprint.   

\subsubsection{Microring resonator}
We next fabricate LNOI microrings with top width varied between 575 and 625 nm. For the ease of phase matching, we make a series of microrings with large radii around 80 $\mu$m (see Fig.~\ref{figure1}(b)), to obtain small free spectrum range (FSR, $\sim$ 2.3 nm for quasi-TM modes). This approach significantly improves the probability of achieving natural phase matching for SHG between two resonant wavelengths, with one around 1550 nm and another around 775 nm \cite{Guo:16}. It worths noting that a larger radius does not lead to any significant improvement on the quality factor as the bending loss is negligible compared to the scattering loss when the radius is larger than 15 $\mu$m. This means that by using a similarly phase matching microring but with 15$\mu$m radius, the SHG efficiency can be improved by over 5 times, simply because of the reduction in the cavity mode volume.  

\begin{figure}[htbp]
\centering
 \subfloat{
   \includegraphics[width=0.45\textwidth]{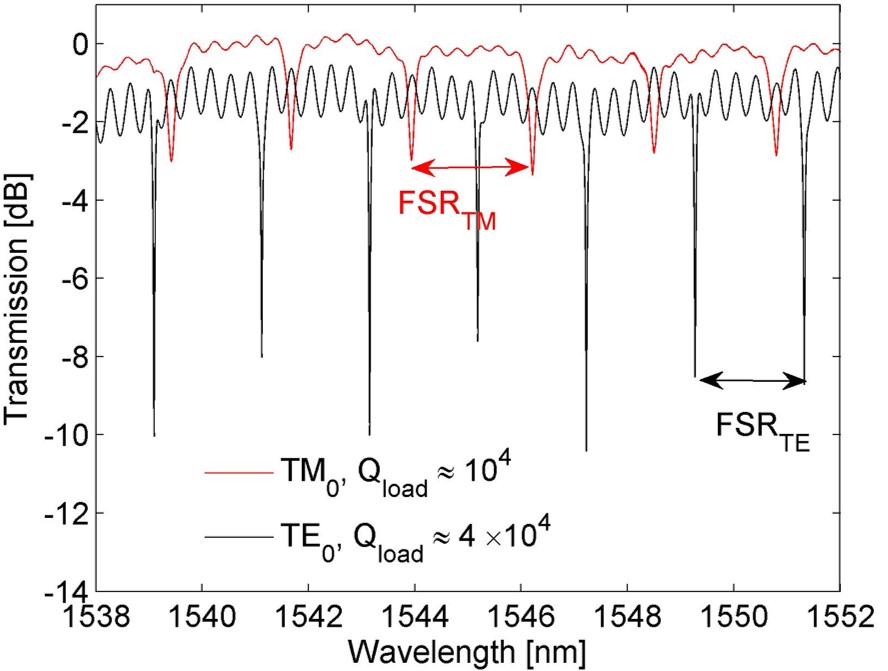}
   \label{fig:sub31}
}
 \qquad
 \subfloat{
   \includegraphics[width=0.46\textwidth]{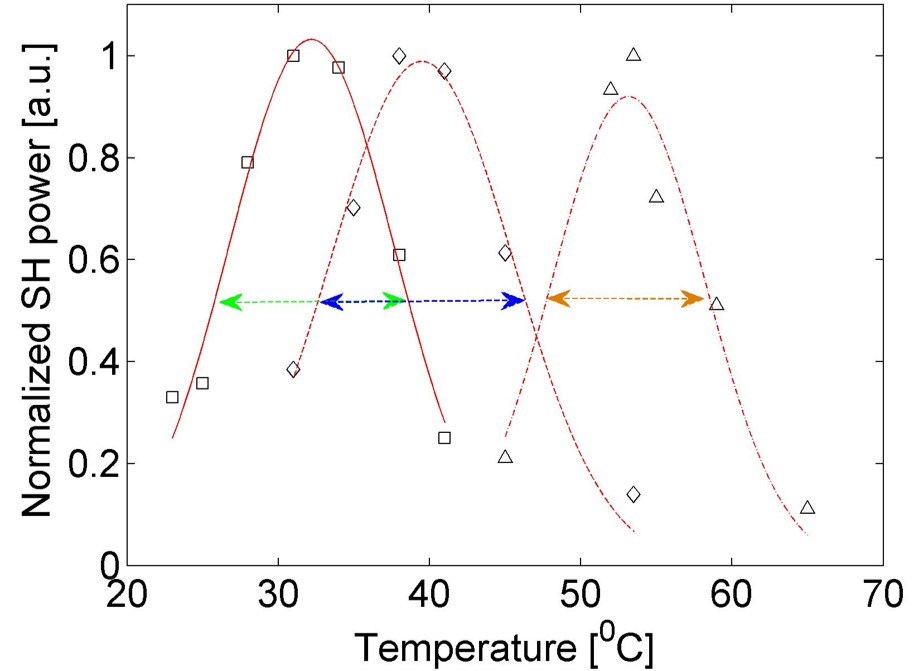}
   \label{fig:sub32}}
 \caption{(a) Transmission spectra of quasi-TE and quasi-TM cavity modes. (b) Temperature dependency of perfect phase matching for different resonances. Square, Diamond, Triangle markers represent resonances around 1544 nm, 1537 nm and 1565 nm, respectively. }
 \label{figure4}
\end{figure}

To characterize the resonant spectra of the microrings, we use an optical spectrum analyzer to measure the transmission of a broadband light beam generated through the amplified spontaneous emission (ASE) of an EDFA. A typical transmission spectrum measurement for an 80-$\mu$m-radius LNOI microring is shown in Fig.~\ref{figure4}(a). In order to fulfill the phase matching condition, we first need to identify the appropriate quasi-TM mode for the pump in telecom wavelength. According to the simulation results, the effective refractive index of a quasi-TE-mode is slightly higher than a quasi-TM mode at the same wavelength. As the FSR is inversely proportional to $n_\mathrm{eff}$, the quasi-TM modes are wider spaced than the quasi-TE modes in the transmission spectrum, which allows us to distinguish the quasi-TM and TE modes. As shown in Fig.~\ref{figure4}(a), the FSR of the quasi-TM and quasi-TE modes are measured to be 2.0 and 2.2 nm, respectively, with the loaded cavity quality factors of $4\times10{^4}$ and $1\times10{^4}$. With the high index contrast and low absorption loss, the quality factor for the current microrings is largely limited by the sidewall roughness. The quasi-TE modes have better mode confinement than the quasi-TM modes, hence experiencing less scattering loss thus a higher intrinsic quality factor. Currently the etching process suffers the surface redeposition and contamination. By optimizing the chamber conditions and fine tuning the etching parameters, it has been shown that the redeposition and contamination can be significantly suppressed, giving rise to a considerably higher quality factor \cite{Zhang:17}. 
The highest SHG efficiency is achieved at critical coupling for both the pump and SH modes while their resonances are perfectly aligned \cite{Guo:16,guosec}. This requires $Q_c$ = $Q_i$, where $Q_c$ and $Q_i$ are the coupling and intrinsic quality factors, and $\Delta \lambda = \lambda_{p}-2\lambda_S$ is zero. As such, high quality factors with narrow resonant linewidths will impose stringent phase matching conditions for SHG. Therefore, an over-coupled microring in the telecom band with broadened resonances offers better spectral overlap between the pump and SH cavity modes \cite{guosec}. Meanwhile, with a lower loaded cavity factor, it is easier to maintain the phase matching without the need for meticulous temperature or electro-optical tuning \cite{natphasemat2010prl}.
 
The phase matching condition for SHG between the quasi-TM cavity modes is satisfied when their azimuthal mode numbers (given by $2\pi n_\mathrm{eff} R/\lambda$) of the SH mode (M2) equals twice that of the pump (M1). This condition is met by accurately engineering the waveguide geometry to match the effective refractive indicates for those modes. In practice, however, the inevitable imperfections in fabrication and optics will result in slight misalignment of the cavity resonances, with $\Delta\lambda\neq 0$ and gives very low SHG efficiency. 

To overcome this difficulty, we apply thermal tuning mechanism to approach the optimum phase matching, by utilizing LN
's different thermo-optic coefficients for those modes, $d_{\lambda}=\frac{d\lambda}{dT}$
. Specifically, we choose the temperature $T$ such that $\Delta \lambda + \Delta d \times T=0$. Here $\Delta d =d_{1550}-2\times d_{775}$, with $d_{1550}$ and $d_{775}$ being the thermo-optic coefficients for the 1550-nm TM$_0$ and 775-nm TM$_2$ modes, respectively \cite{Guo:16,guosec}. For the current devices, however, $\Delta d$ is relatively small so that the amount of resonances misalignment that can be compensated is fairly limited. Other measures for phase matching engineering, including those by periodic poling or electro-optical modulating, may be desirable, especially for the microrings with very small radii. For the present, we use large microrings to increase the number of resonant modes within the phase matching window (see Fig.~\ref{figure4}(a)), so that $\Delta \lambda$ of several resonances can be smaller than $(\Delta d \times T)$ in the same microring. This improves the chance of achieving natural phase matching and help the tolerance of the fabrication and optics imperfections. In Figure~\ref{figure4}(b), we measure the temperature dependency of SH power for three different resonances measured on the same chip. The full-width at half-maximum (FWHM) of the fitted Lorentzian shape curves are $12.5\pm1.5\, ^{\circ}$C, which agrees with the theoretical value of $11.0\,^{\circ}$C (see supplementary for details).

\begin{figure}[htbp]
\includegraphics[width=5in]{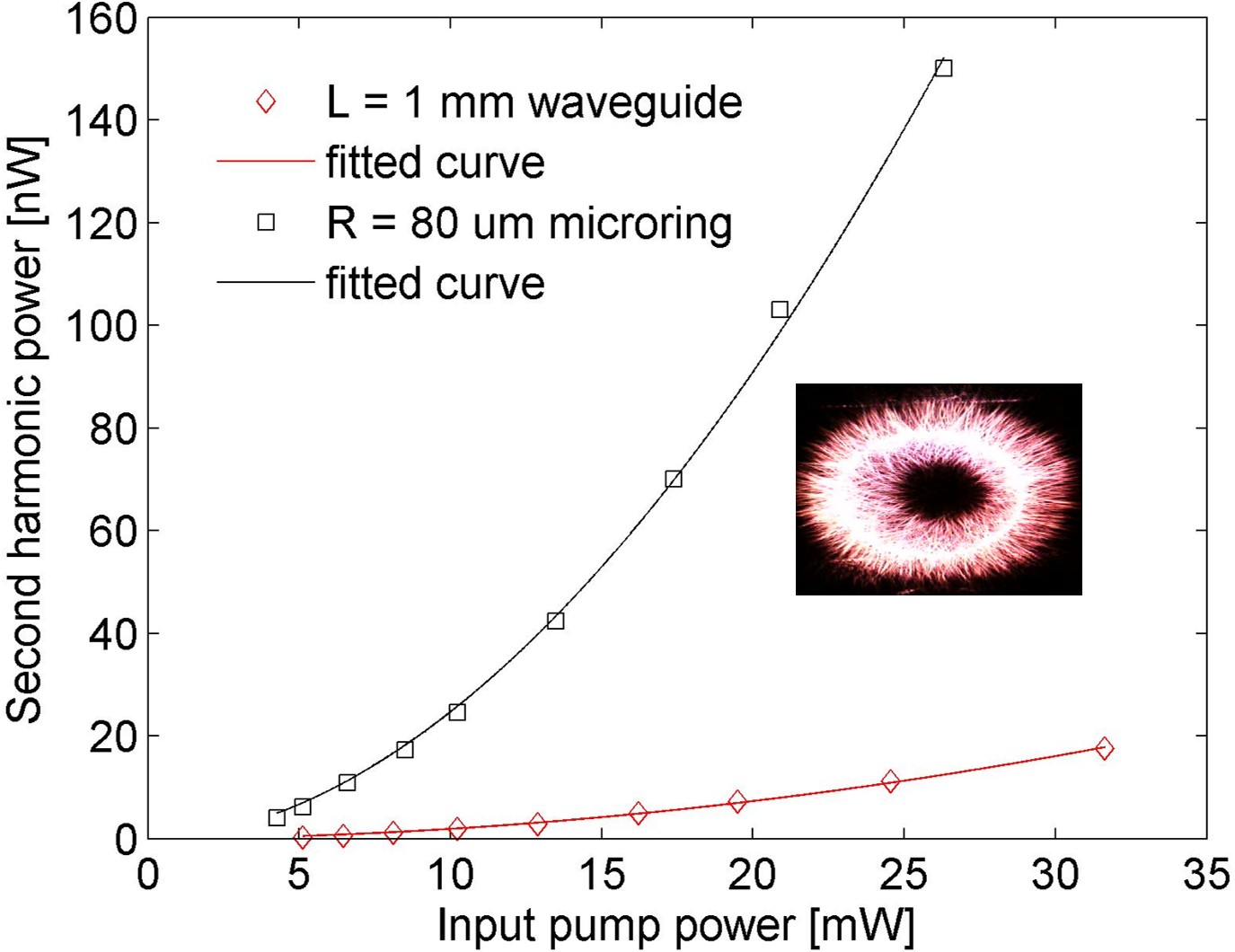} 
  \centering
\caption{Comparison of second harmonic generation between a straight waveguide and a microring resonator of similar cross-section geometry. Inset: The microring illuminated by the bright SH light generated inside with few-milliwatts input pump power.}
\label{figure5}
\end{figure}


We measure the SH power generated in a 1-mm straight waveguide and an 80-$\mu$m microring with different pump powers. For a fair comparison, both devices have input and output couplers with similar dimensions, so that the fluctuation of the measured coupling efficiencies is within 1 dB. As shown in Fig.~\ref{figure5}, the generated SH power exhibits a quadratic dependency of the pump power in both the waveguide (red line) and microring (black line), each with a fitted polynomial power of  1.95 and 1.88, respectively. This indicates SHG in the un-depleted pump regime for both devices \cite{Guo:16}. For the same pump power, the out-coupled SH power from the microring is 10 times higher than the straight waveguide, despite the present cavity is in the under-coupled region for the SH TM$_2$ mode. According to the simulation results, about 75 percent of the generated SH light is lost in the microring. This can be improved by two times through utilizing additional wrap waveguide\cite{Guo:16} to efficiently out-couple the SH light without affecting quality factor of the cavity modes in telecom band. Besides, we estimated that given the current intrinsic quality factor, the internal nonlinear conversion efficiency can be further enhanced by 10 times via critical coupling scheme for the pump and using 15 $\mu$m radius, $\sim \eta_{internal} \propto \frac{Q_c^2}{R\,(Q_c+Q_i)^4}$ \cite{Guo:16}. By using compact microring structure, we could alleviate the non-uniformity and fabrication errors present in the long waveguide case. In addition, dual resonant modes matching in microring resonator ensures an well defined phase matching bandwidth with a single principal peak while suppressing those secondary side peaks observed in the straight waveguide. This can be an important feature for all- optical classical or quantum information processing on-chip\cite{PhysRevX.5.041017,Shahverdi2017}.

\section{Conclusion}
We have demonstrated a modal engineering approach to attaining natural phase matching for second-order nonlinear processes in both straight waveguides and microring resonators on a monolithic Z-cut LNOI via E-beam lithography and top down dry etching method. In sub-mircron waveguides, we observe phase-matched SHG with $26 \%W^{-1}cm^{-2}$ efficiency. In microring resonators of similar cross-section geometry, we also achieve naturally phase matched SHG with a net enhancement of 10 times in efficiency. As the phase matching occurs within the quasi-TM modes, we gain access to LN's highest nonlinear susceptibility tensor element (d$_{33}$) without restriction in the waveguiding direction. 
Achieving submicron spatial confinement of light and empowered by the highest nonlinear coefficient lithium niobate, this work may open a new horizon for high complexity, dense nonlinear photonic integrated circuit. Thus, it may extend a plethora of nonlinear optics applications on-chip, including optical harmonic generation, frequency mixing, octaves spanning frequency comb, interaction free optical switching, and photon-photon interaction for classical and quantum information processing.

\begin{acknowledgement}

This research was support in part by the National Science Foundation (Award No. ECCS-1521424 and No. EFRI-1641094). Fabrication was performed at CUNY Advanced Science Research Center Nano Fabrication Facility. 
\end{acknowledgement}

\begin{suppinfo}

\end{suppinfo}

\bibliography{achemso-demo}

\end{document}


\begin{suppinfo}

The experimental setup for signal depletion measurement is shown in S.~\ref{schematic}. The pump is a 10-MHz pulse train with 130-ps full width at half maximum (FWHM), created by modulating a narrow-band CW laser using electro-optical modulator (EOM). To generate the required high peak power pump pulses ($\sim$ 1 W), we amplify the optical pulses by using a two-stages erbium-doped fiber amplifier (EDFA) system. The quasi-CW signal with wider pulse width ($\sim$ 10 ns FWHM) is synchronized and temporally aligned with the pump pulse. By using short pump pulses and quasi-CW signal pulses, we reduce the required average power coupled into the waveguide hence avoiding the photo-refractive damage to the sample \cite{moore2016efficient}. The polarization of both pump and signal pulses are independently adjusted by using fiber polarization controllers, before being combined into the device through a dense wavelength-division multiplexers (DWDM).
\captionsetup[figure]{labelfont={bf},labelformat={default},labelsep=period,name={S.}}
\begin{figure}[htbp]
\includegraphics[width=5in]{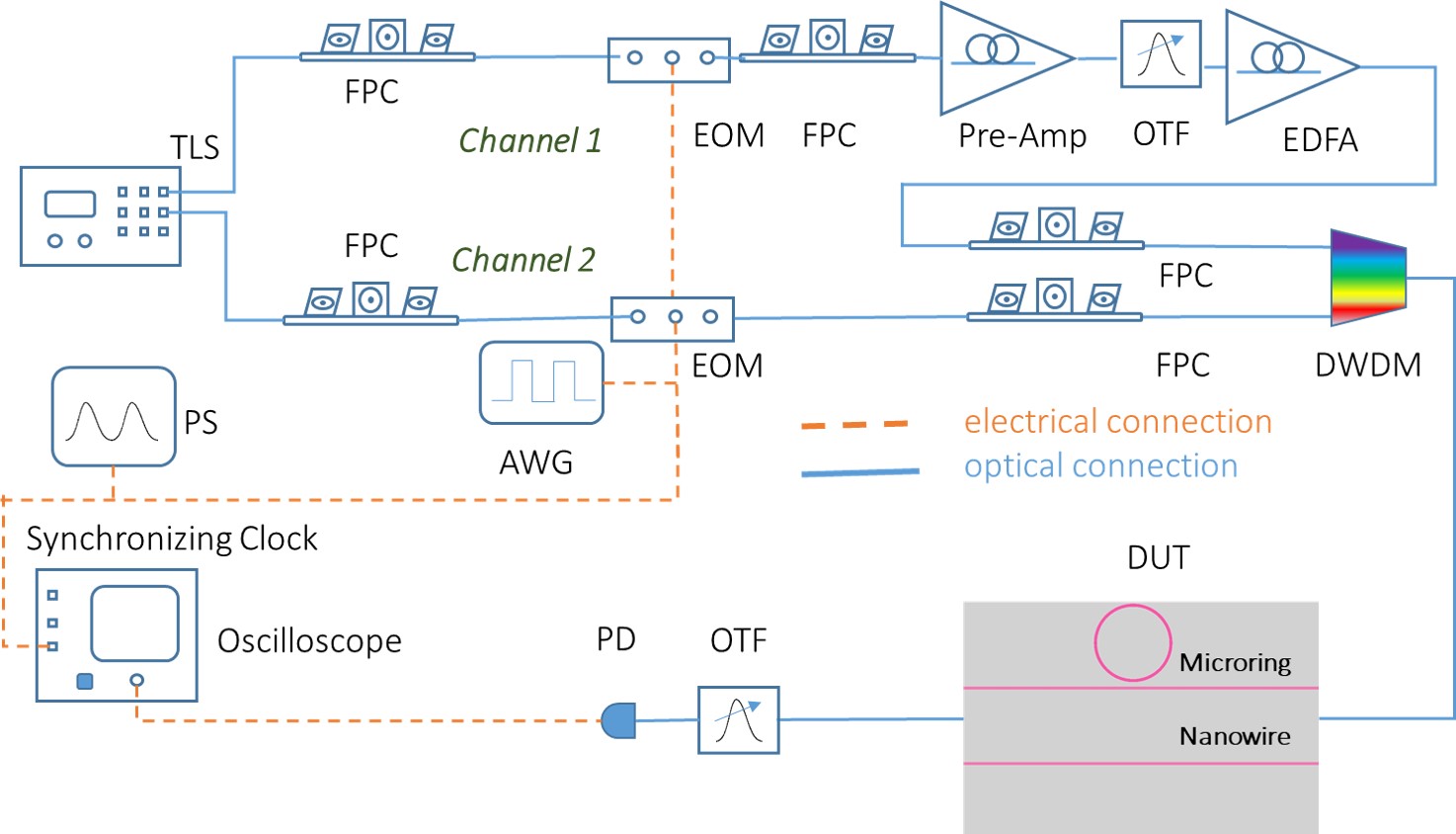} 
  \centering
\caption{Experimental setup with two synchronized optical channels. Channel 1 creates a pump pulse train with 10 MHz repetition rate and 130 ps FWHM. Channel 2 produces a quasi-CW signal with 10-MHz repetition rate and 10-ns FWHM. TLS, multichannel narrow linewidth ($<$100 KHz) tunable laser system, PS, electrical pulse generation system, EOM, electro-optic modulator, FPC, fiber polarization controller, AWG, arbitrary waveform generator, DWDM, dense wavelength-division multiplexer, TOD, tunable optical delay, OTF, optical tunable filter, DUT, device under test.}
\label{schematic}
\end{figure}

By exploiting modal phase matching between 1550-nm TM$_0$ and 775-nm TM$_2$, we have utilized the largest nonlinear coefficient with diagonal tensor $d_{33}$, $\sim$ 27 pm/V. The simulated mode profiles for 1550-nm TM$_0$ and 775-nm TM$_2$ are shown in the inset in Fig.1 (a). The SH conversion efficiency of a LN waveguide is given as, \cite{hiroshi09}
 \begin{equation}
   \eta_{norm} = \frac{8\pi^2}{\epsilon_0 c \lambda^2_{2\omega}} \,  \frac{d_{33}^2}{n^{2\omega}_{eff} (n^{\omega}_{eff})^2} \, \frac{\iint\,E_{2\omega}^{*}\,E_{\omega}^2\,dy\,dz}{\sqrt{\iint\,|E_{2\omega}|^2\,dy\,dz} \,\iint\,|E_{\omega}|^2\,dy\,dz}\,sinc^2(\Delta K L/2),   
   \label{eq1}
 \end{equation}
where $c$ is the speed of light in vacuum, $\epsilon_0 $ is the vacuum permittivity, $d_{33}$ is the effective nonlinear coefficient of lithium niobate, $\Delta K$ is the wave vector mismatch, $n^{(\omega,2\omega)}_{eff}$ and $E_{(\omega,2\omega)}(y,z)$ are the effective indexes and electrical fields of 1550-nm TM$_0$ and 775-nm TM$_2$ modes, respectively. Here we consider x-axis as the propagation direction but not restrict to it since the effective refractive index for quasi-TM mode is equivalent over the entire x-y plane. The theoretical normalized efficiency is calculated to be approximately $61 \%W^{-1}cm^{-2}$ for a perfectly phase matched ($\Delta K = 0$) LN waveguide. One can see that the efficiency is limited by those integral terms which is the effective transverse mode-overlapping factor between 1550-nm TM$_0$ and 775-nm TM$_2$ modes in the submicron waveguide, an inevitable drawback of our modal phase matching scheme despite obtaining a plausible conversion efficiency. Thus, quasi-phase matching by periodic poling \cite{Chang:16} will be an important immediate future work to improve the conversion efficiency to the next level for many applications.

\captionsetup[figure]{labelfont={bf},labelformat={default},labelsep=period,name={S.}}
\begin{figure}[htbp]
\includegraphics[width=5in]{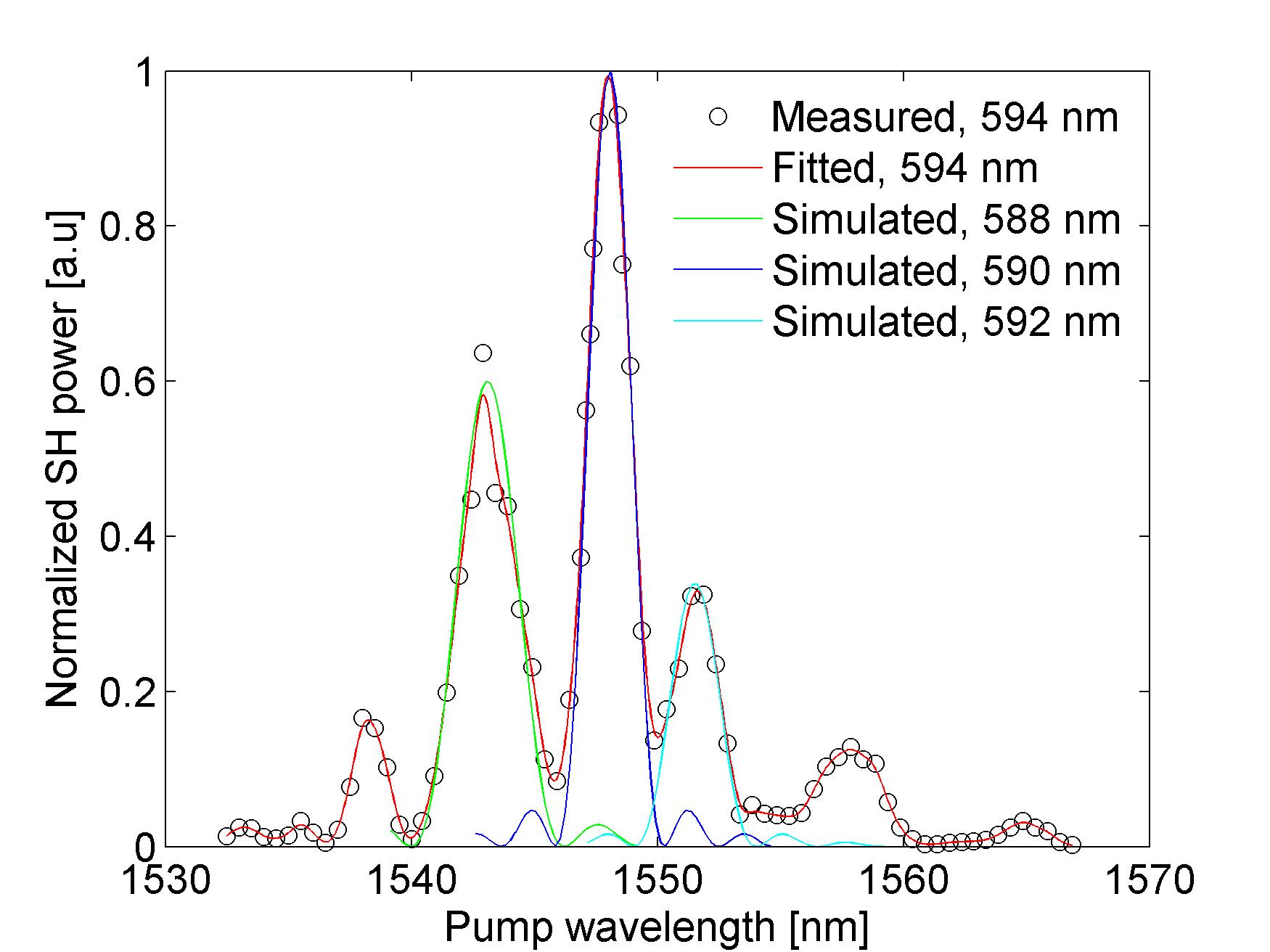} 
  \centering
\caption{Measured and simulated phase matching curve in telecom band for  L = 1 mm LN waveguide. Dimension of the fabricated waveguide is 400 nm thickness, 594 nm top width and $70^\circ$ sidewall angle, working temperature 28.5$^{\circ}$C.}
\label{s2}
\end{figure}
As shown in S.~\ref{s2}, we observe multiple side peaks along with a central phase matching peak on this 1-mm long waveguide. Ideally, a perfectly phase matched and uniform waveguide is desirable to attain maximum SH efficiency with a well defined principal peak as described by Eq.~\ref{eq1}. However, small variations in physical properties along the waveguide including its width and thickness, side wall angle, refractive index or even the material compositional inhomogeneity will induce optical inhomogeneity\cite{Nash1970}, which in turn cause a small net phase shift on the 1550-nm TM$_0$ and 775-nm TM$_2$ lightwaves propagating inside it\cite{Wang:17}. Phase matched 1550-nm TM$_0$ and 775-nm TM$_2$ modes mean infinitesimal wave vector mismatch ($\Delta K \approx 0$) despite non-uniformity over the waveguide, thus allowing long coherent buildup length ($\gg$ 1 mm)\cite{Boyd200869,limfejer1990}. In order to shed light on the measurement result, we establish a empirical model to reflect the physical variation on the entire 1-mm long waveguide by decomposing it as 3 sections of shorter waveguide with different top-width and effective length. The spectral dependency of normalized SH power for waveguide with top width of 588 nm, 590 nm. 592 nm are plotted in S.~\ref{s2}. Interestingly, only a few nm difference in the waveguide's top width is sufficient to shift the phase matched wavelength thus causing the presence secondary phase matching peaks that matches experimental observation. The simulation result suggests that the physical variation on the waveguide is roughly equivalent to several sections of waveguide with different phase matching wavelengths. This also indicates the phase matching condition can be much altered by optical inhomogeneities in the waveguide arise from the fabrication process. 

Since slight variation of the refractive index and waveguide dimensions will affect the phase matching profoundly, we also investigated the influence of device temperature on the phase matched LN waveguide. For simulation, we first use the temperature-dependent generalized Sellmeier equation~\cite{PhysRevB.48.15613} to compute the refractive indexes of ordinary and extraordinary axis at different temperatures. Then we model the waveguide with temperature corrected refraction indexes and plot the corresponding phase matching curve as shown in S.~\ref{sup}. One can see that the phase matching wavelength is red shifted with increasing temperature at the rate of 0.062 nm/$^{\circ}$C. To ratify the simulation result, we measure the phase matching curves at 28.5 $^{\circ}$C and 48.5 $^{\circ}$C and obtain the shift of phase matching wavelength. To avoid pump power induce dynamical thermal effect\cite{Carmon:04}, We use a narrow linewidth CW laser at low power (un-depleted pump) region and measure the SH power as a function of wavelength by using the "Max Hold" function on the OSA, as shown in S.\ref{sup} (b). We notice that the primary phase matching peak is red shifted by 0.13 nm with temperature increased by 20 $^{\circ}$C, which is corresponding to 0.065 nm/$^{\circ}$C. The measured wavelength shift agrees well with our theoretical simulation results, proving that thermo-optics effect is the dominant factor on the temperature-dependency of phase matching wavelength for our device.
\captionsetup[figure]{labelfont={bf},labelformat={default},labelsep=period,name={S.}}
\begin{figure}[htbp]
\centering
 \subfloat{
   \includegraphics[width=0.45\textwidth]{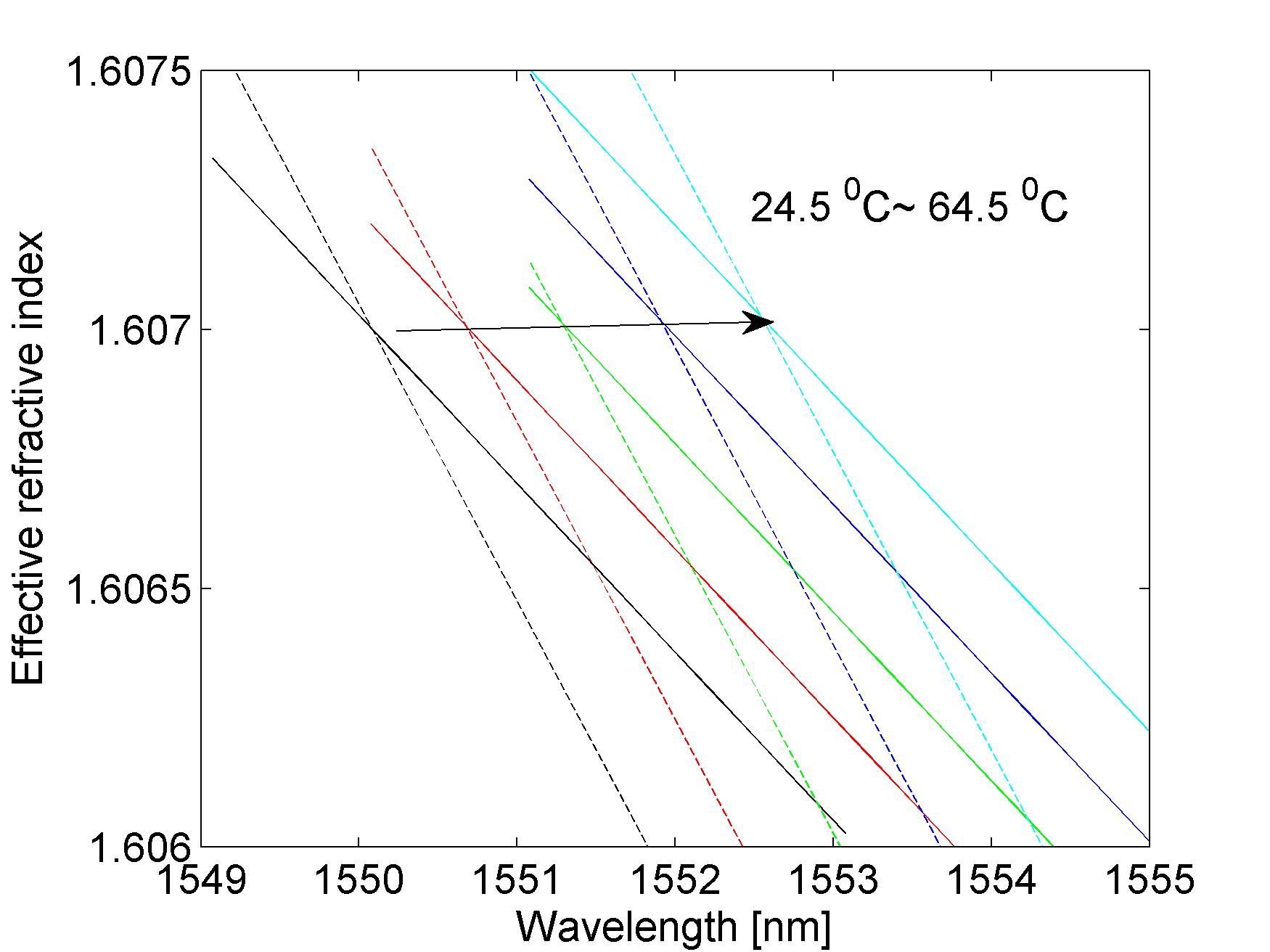}
   \label{fig:sub41}
 }
 \qquad
 \subfloat{
   \includegraphics[width=0.45\textwidth]{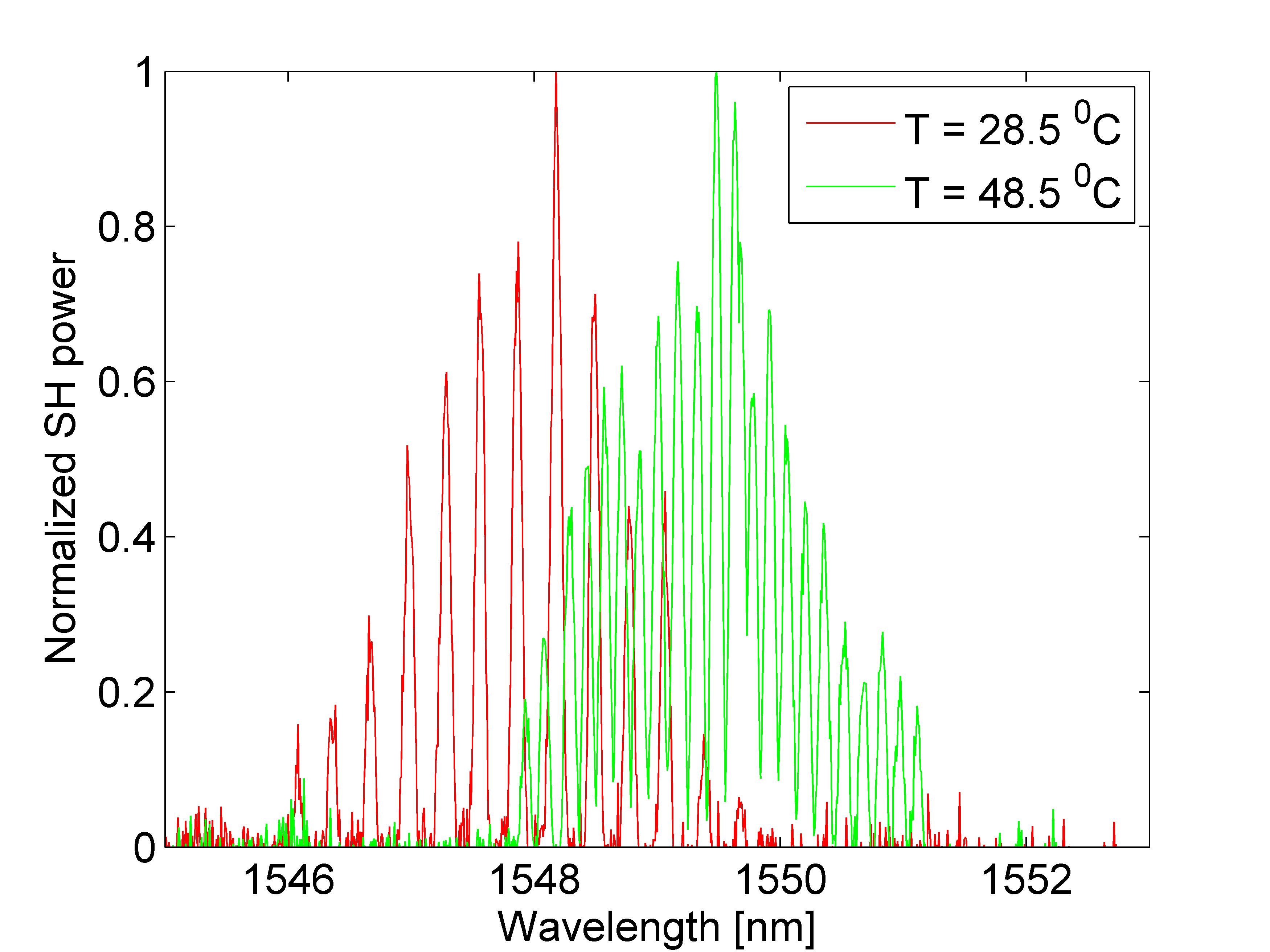}
   \label{fig:sub42}
 }
 \caption{(a) Simulated temperature dependency of phase matching central wavelength. Solid and dash line represent 1550-nm TM$_0$ and 775-nm TM$_2$, respectively. Black, red, green, blue and cyan colors represent different temperatures from 24.5$^{\circ}$C to 64.5$^{\circ}$C with 10 $^{\circ}$C step. (b) Measured phase matching curve at different temperatures. Here the individual peaks are due to the max-hold function on optical spectrum analyzer.}
 \label{sup}
\end{figure}
As the phase matching in microring resonator can only be fulfilled by the means of fundamental and SH resonant modes matching, we first compute azimuthal mode number to obtain the  M = 521 ( 1042) for fundamental (SH wavelength), where $n^{\lambda}_{eff} = 1.606777 (1.606765)$ is extracted at T = 24.5 $^{\circ}$C. Then, we can compute the temperature dependence resonant wavelength $\lambda_M(T)$ for same azimuthal mode in both fundamental (M = 521) and SH (M = 1042) wavelength by using temperature-dependent generalized Sellmeier equation to extract  $n^{\lambda}_{eff}(T)$. Subsequently, we obtain the thermo-optic coefficients $d_{\lambda_M} = \frac{d\lambda_M}{dT}$ for 1550-nm TM$_0$ and 775-nm TM$_2$ cavity modes to be 0.0193\,nm/$^{\circ}$C and  0.0167\,nm/$^{\circ}$C, respectively. The difference of the thermo-optic coefficient between cavity resonances at fundamental and SH wavelength, $\Delta d = d_{1550} - 2\times d_{775} $= -0.0141 nm/$^{\circ}$C, is a crucial degree of freedom for compensating the small resonant modes mismatch by $\Delta \lambda_{eff} = \Delta \lambda + \Delta d \times T$ \cite{Guo:16,guosec}. Refering to the equation \cite{Guo:16} $\Delta T = \frac{\lambda_{1550}}{Q_{775}\,|\Delta d|}$, we estimate that $\Delta T$ = 11.0\,$^{\circ}$C (The loaded quality factor of 775-nm TM$_2$ cavity mode is approximately to be $10^4$ since the microring is under-coupled at 775-nm according to simulation results). The estimated $\Delta d$ for LNOI microring is 10 times larger than other comparable platform such as aluminum nitride microring \cite{Guo:16}, which implies it has more stringent phase matching condition than the latter even with identical quality factor. Instability of linear and nonlinear properties of microresonator structre on LNOI due to dynamical thermal behavior will be an issue that need to be carefully addressed. When Q factor is approaching $10^5$ ($10^7$), using a stable $\pm 0.1\,^{\circ}C$ ($\pm 0.001\,^{\circ}C$) temperature controlling system, along with delicate electrical tuning or locking mechanism on chip will be necessary.
\end{suppinfo}

\bibliography{achemso-demo}